# INTERROGATION OF A BUBBLE IN THE INDIAN MARKET

**Ganapathy G Gangadharan**
Final Year Student, Ramaiah University of Applied Sciences, Bangalore, India

**\*Dr N. Suresh**
Professor, Faculty of Management and Commerce,
Ramaiah University of Applied Sciences, Bangalore, India
*Corresponding Author: nsuresh.ms.mc@msruas.ac.in

**ABSTRACT**

*Emerging markets such as India provide the investors with returns far greater than those in developed markets; taking the average returns from the period 1995 to 2014 the returns are 4.714% to 3.276% of the developed market (US not included). Majority of emerging markets commenced joining with the capital market of the world, thus allowing huge inflow of capital which in turn paved the path for economic growth. Even though the emerging markets provide high returns these may also be an indication of a bubble formation. Detection of a bubble is a tedious task primarily due to the fundamental value of the security being uncertain, the randomness of the fundamentals of the market make detecting bubbles an arduous task. Ratios that foretold the financial crisis of 2007- Market Capitalization to GDP (Buffet Indicator), Price to Earnings Ratio (PE Ratio), Price to Book Value (PB Ratio), Tobin's Q. Data is collected from the 1999-2000 from various Indian indices such as NIFTY 50, NIFTY NEXT 50, NIFTY BANK, NIFTY 500 S&P BSE SENSEX, S&P BSE 100. The paper utilizes the ratios mentioned above to detect and back track various bubble episodes in the Indian market; methodology used is the Philips et al (2015) right tailed unit test. The paper is also inclined to take steps to mitigate the effects of bubble by amending the financial policies and the monetary liquidity of the financial system.*

**Cite this Article:** Ganapathy G Gangadharan, Dr N. Suresh, Interrogation of a Bubble in the Indian Market, *Journal of Management*, 6 (3), 2019, pp. 64–70.
http://www.iaeme.com/JOM/issues.asp?JType=JOM&VType=6&IType=3

## 1. INTRODUCTION

The Indian economy commenced trading electronically since 1995, this introduced the collection of tick data which after thorough analysis can provide insight whether the Efficient Market Hypothesis (EMH) is coherent (Dutta, Gahan, & Panda, 2016). The EMH purports that stocks reflect information and investors decisions are postulated of these information, thus negating any individual to time the market correctly to buy undervalued and sell overvalued (Nagpal, Jain, 2018). The notion that investments are all fairly priced may not necessarily be true. The herd mentality of the investors which is quite evident during a bull or a bear rally in which there is optimism in the market or panic is the reality (Dutta, Gahan & Panda, 2016).

Majority of emerging markets commenced joining with the capital market of the world, thus allowing huge inflow of capital which in turn paved the path for economic growth (Bekaert and Harvey, 2000). Albeit such inflow at such amount may inevitably lead us to the bubble (Kim and Yang, 2009). Even though the emerging markets provide high returns these may also be an indication of a bubble formation. Uncertainty of the fundamentals are one of the main causes of bubble formation and the emerging markets are highly susceptible to such formations (Parke and Waters, 2007).

Although Price-dividend ratios are crucial in bubble identification and used in most of the bubble detection models (Almudhaf 2017; Chen and Xie 2017; Su et al. 2018) there are other ratios that can provide a deeper insight in successfully detecting a bubble. Ratios that foretold the financial crisis of 2007- Market Capitalization to GDP (Buffet Indicator), Price to Earnings Ratio (PE Ratio), Price to Book Value (PB Ratio), Tobin's Q (Kuepper, 2014). There is evidence that indicate PE ratio does claim dominance over dividend yield in valuation models(Barker,1999). In this paper the discussion pans out on the utilization of a few ratios mentioned above in the Indian Market Indices; the methodology of calibrating and detection of a bubble is done through Philip et al. (2015) method- Generalized supernum Augmented Dickey Fuller Method (GSADF), which is consistent in multiple bubble detection (Chen and Xie, 2017).

## 2. LITERATURE REVIEW

Several tests were conducted to check the efficiency of the Indian market. For instance, Run test, Variance Ratio test, ARIMA (Auto-regressive Integrated Moving Average), Auto Regression (AR), Moving Average (MA) indicate that the investors are often oblivious to the risk during the bubble and quite perplexed post bubble since they tend to overreact to the bear and bull cycles (Siddiqui and Nabeel, 2013).Various studies in different markets have shown us that they don't possess tools that can deal with pre ante bubble economically, financially and monetarily. Thus, causing the scenario of 1997 South-east Asia wherein the burst impacted the financial and social sectors (Froot and Obstfeld 1991; Hommes et al. 2005). There are analysts who opine that the reason for the incompetent manner of addressing such developments in the capital and real estate market are due to the vested interest of the milieus of finance and banking who generally profit from the bubble period. There countries where the banks are working with almost no supervision or necessary restrictions, ergo fortifying the inflation of such events.

It is quite convenient to have a hindsight 20/20 post-bubble whereas it is arduous to increase one's wealth by timing it (Ball, 2009). There are methods to detect such indicators which can aid in detecting bubbles and keep its development under supervision in order to curb the financial and monetary impacts of it by economic policies (Girdzijauskas et al, 2009). The methodology is the right tailed unit test of Philips et al (2015) which is one of the effective tools to detect multiple bubbles and since the emerging markets are characterized by volatility compared to the developed economies market (Barkoulas et al., 2000; Kasman et al, 2009).

The paper by Chang et al (2014) performed the recursive unit root test of Philips et al (2011) and Philips et al (2013) in the BRICS nations in order to detect bubbles and they've been successful in doing so. They have used Price to dividend ratio, this paper is inclined to use ratios such as PE ratio, PB ratio, Buffet Indicator, Tobin's Q etc. in various sectors and indices of the Indian stock market. The paper aims to ascertain which ratios aid efficiently in detecting bubble in various Indian indices based on the following:
- Benchmark indices-S&P BSE Sensex, NSE Nifty 50, NSE NIFTY NEXT 50
- Sectoral indices-NIFTY BANK
- Broad-market indices-NIFTY 500 and S&P BSE 100

**Table 1** The SADF and GSADF results for the Price-earnings ratios of Indian Indices

| 1999-2000 | Test critical values*: | GSADF | Prob. | Test critical values*: | SADF | Prob. |
|---|---|---|---|---|---|---|
| NIFTY 50 | | 3.728618** | 0.0000 | | 3.728618** | 0.0000 |
| 99% level | 2.773104 | | | 1.968328 | | |
| 95% level | 2.165766 | | | 1.419215 | | |
| 90% level | 1.924111 | | | 1.161121 | | |
| NIFTY NEXT 50 | | 3.728618** | 0.0000 | | 0.877954*** | 0.0290 |
| 99% level | 1.842877 | | | 1.161273 | | |
| 95% level | 1.096555 | | | 0.700554 | | |
| 90% level | 0.816866 | | | 0.457592 | | |
| NIFTY BANK | | 0.653972*** | 0.4920 | | -0.398567 | 0.6350 |
| 99% level | 2.776593 | | | 2.214867 | | |
| 95% level | 1.908310 | | | 1.224751 | | |
| 90% level | 1.564568 | | | 0.918405 | | |
| S&P BSE SENSEX | | 0.148245*** | 0.4640 | | -1.128701 | 0.7910 |
| 99% level | 3.763064 | | | 3.013991 | | |
| 95% level | 2.004090 | | | 1.285283 | | |
| 90% level | 1.338134 | | | 0.854516 | | |
| S&P BSE 100 | | 2.484461*** | 0.0290 | | 1.988333*** | 0.0230 |
| 99% level | 3.763064 | | | 3.013991 | | |
| 95% level | 2.004090 | | | 1.285283 | | |
| 90% level | 1.338134 | | | 0.854516 | | |
| NIFTY 500 | | 2.858897** | 0.0010 | | 3.004325** | 0.0010 |
| 99% level | 2.064438 | | | 1.968328 | | |
| 95% level | 1.577620 | | | 1.419215 | | |
| 90% level | 1.269780 | | | 1.161121 | | |
| * -Critical values are obtained from Monte Carlo simulation(1000 replication) | | | | | | |
| ** -1% Significance level | | | | | | |
| ***-5% Significance level | | | | | | |

## 3. METHODOLOGY

The daily and monthly price to earnings, price to book value and dividend yield is taken from NSE and BSE sites with various periods based on the occurrence of a Bubble and the availability of the data pertaining to those specific market. We have covered six market indices (NIFTY 50, NIFTY NEXT 50, NIFTY BANK, BSE 500, BSE SENSEX, BSE BANK, NIFTY BANK). The index chosen represents two of the leading stock exchanges in India which have financially sound and well trenched companies. The BSE alone has companies with a combined market cap of 1.43 trillion dollars (USD). Hence reliable data and the ones that ought to react during any exuberance period, has been taken for bubble detection. The following equation is what the test is based on:

$$x_t = \alpha + \delta x_{t-1} + \sum_{i=1}^{P} \emptyset_i \Delta x_{t-j} + \varepsilon_t \qquad (1)$$

The level index is indicated by $X_t$, the intercept by α, the maximum number of lags by p and the ε depicts the white noise error term. The null hypothesis is that there is a unit root i.e. $\delta=1$ and the alternate hypothesis is that there is an exuberance period i.e. $\delta >1$ indicating presence of bubble (based on the right tailed Augmented Dickey Fuller test).

The model based on the Philips et al (2015) is capable of including multiple bubbles:

$$X_t = (X_{t-1} + \varepsilon_t)\,1\{t\epsilon N_0\} + (\delta_T X_{t-1} + \varepsilon_t)\,1\{t\epsilon F_i\}$$
$$+ \sum_{i=1}^{L}\left(\sum_{k=\tau_{ib}}^{t}\varepsilon_k + X^*_{\tau_{ib}}\right)1\{t\epsilon N_i\}.$$

L denotes the bubble sample and each interval fraction.

$$F_i = [\tau_{ie}, \tau_{ib}] \text{ for } i = 1,2,\ldots,L.$$

Provided $X^*_{\tau_{ib}} = X_{\tau_{ie}} + X^*_i$, where

$$X^*_i = O_p(1) \text{ for all } i \text{ and subperiods of pure random walk as } N_0 = [1,\tau_{1e}), N_j = (\tau_{j-1b}, \tau_{je}) \text{ with } j = 1,\ldots,L-1 \text{ and } N_K = (\tau_{Lb}, T].$$

The Philips et al (2015) model is performed often on the string of data windows and backwards SADF test on which the performance inference is based on as follows:

$$BSDF_r(r_0) = \sup_{r1\epsilon[0, r2-r0], r2=r}\{DF^{r2}_{r1}\}$$

Where r1 is the initial point and r2 is the final point of the regression. DF stands for the Dickey Fuller test.

Bubble detection and Date-stamping

The methodology adopted is the Philips et al. (2015) which works on the assumption of the random nature of the data. This enables the model to detect formation and collapse of bubble in the market. Thus, the date stamping and identifying bubble in the stock market is achieved using the Generalized Supernum Augmented Dickey Fuller test (GSADF).

The null hypothesis is that there exists a unit root and the alternative being there exists a bubble i.e. there is no unit root. The Philips et al. (2015) has the capability to include multiple bubble and performs often the recursive right tailed test on a series of data windows and performs inference based on the backwards Supernum Augmented Dickey Fuller (SADF).

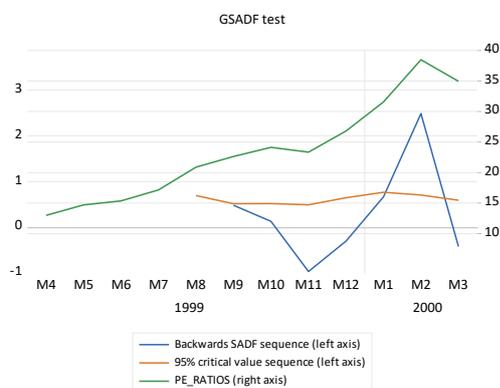

S&P BSE 100

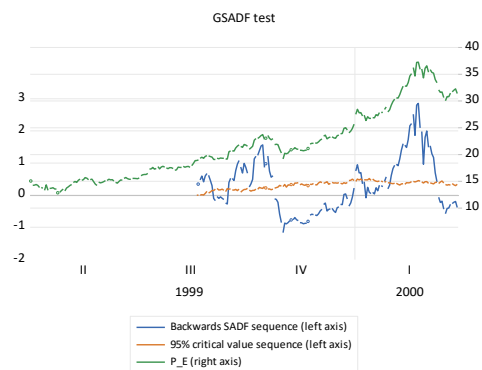

NIFTY 500

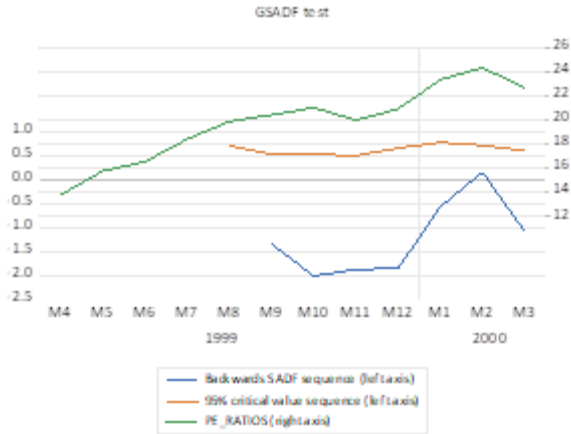
S&P BSE SENSEX

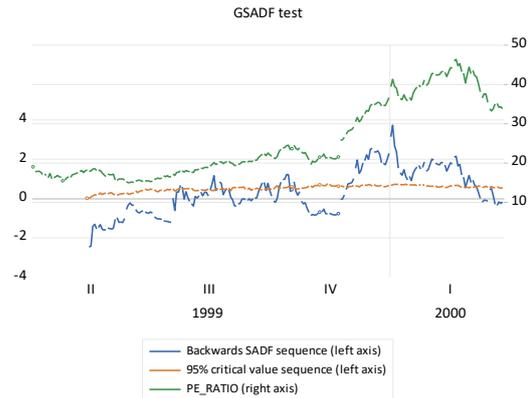
NSE NIFTY 50

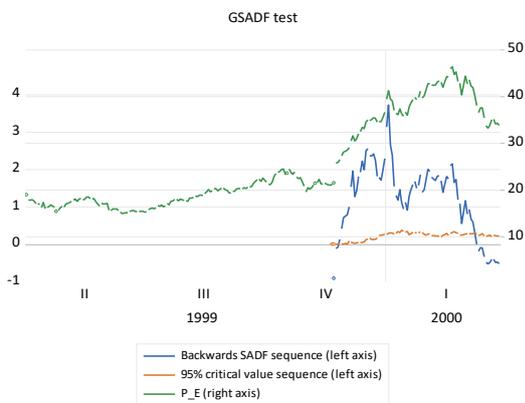
NSE NIFTY NEXT 50

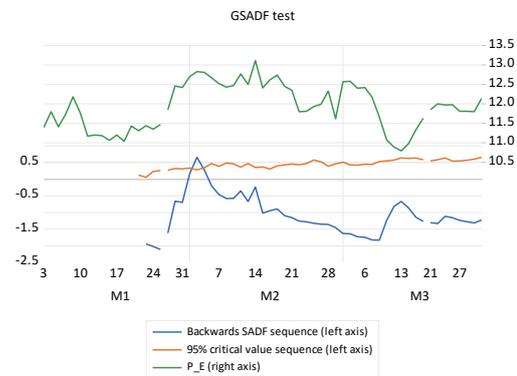
NSE   NIFTY BANK

**Figure 1**. GSADF of pe ratios of the indian indices during the dot com bubble (1999-2000)

## 4. RESULT

Table 1 indicates the Supernum Augmented Dickey Fuller test (SADF) and Generalized Supernum Augmented Dickey Fuller test (GSADF) result for single bubble and multiple bubbles for the phenomenon known as the dot com bubble. The critical values are a result of the Monte Carlo simulation (1000 replications). The lag order is set to zero. The right tailed critical value of GSADF exceeds that of NIFTY 50, NIFTY NEXT 50, NIFTY 500 and S&P BSE 100 as indicated in Table 1. The PE ratios of these indices are placed in the historic exuberance period- the dot com bubble of 1999-2000. The PE ratios of these indices during these periods indicate exuberance which can be inferred as the existence of the bubble. The period is characterized by high deviation from the fundamental values and is further encouraged by the speculation and herd mentality- sentiment or overconfidence. That being said, the GSADF has been scoring lower than the critical values of S&P BSE SENSEX and NIFTY BANK. This can be a suggestion that these indices never experienced a bubble period during 1999- 2000 phenomenon.

This is further preceded by date stamping the inception and burst of the bubble in each Indian Index. Figure 1 sheds light on the exuberance period of the NIFTY 50, NIFTY NEXT 50, S&P BSE SENSEX, NIFTY BANK, NIFTY 500 and S&P BSE 100.

The Fig. 1 shows us the bubble periods during 1999-2000 for various Indian indices and further indicates the exuberance occurrence in S&P BSE 100, NSE NIFTY 500, NIFTY 50,

NIFTY NEXT 50. The S&P BSE 100 has a spike during the beginning months of 2000, the spike lasts for two months January and February (M1 & M2) and falls off by the start of the third month. Thereby exhibiting the exuberance during the dot com bubble.

The NIFTY 500 on the other hand during the same time has a spike of exuberance between the 3$^{rd}$ and 4$^{th}$ quarter of 1999 later on another minute spike from the end of the last quarter of 1999 to beginning of 2000. Finally, a massive surge in almost the end of the 4$^{th}$ quarter of 1999 to almost the 1/3$^{rd}$ of the 1$^{st}$ quarter of the year 2000.

The NIFTY 50 too indicates similar level of exuberance for the entire period of the dot com bubble (1999-2000). The major spikes during the last quarter of 1999 and subsiding before the end of the first quarter of 2000. The tale was quite different for NSE NIFTY NEXT 50 the surge in the exuberance lasted for almost two quarters- commencing in the start of the last quarter 1999 and subsiding before the half of the 1$^{st}$ quarter of 2000. It is quite evident that the Indian Indices experienced multiple bubbles during the dot com period.

An environment where irrationality and sentiment can dominate the rational of an investor, these are sure signs and thereby supplements the possibility of a bubble. The findings of the various exuberance periods of Indian Indices does bear with the above-mentioned rationale of the behavioral finance theories and various credit policies of the time.

## 5. CONCLUSION

The Indian market like other emerging markets have allured investors for its growth and quite satiable returns. The methodology unit root test by Philips et al (2015) has aided to detecting multiple bubbles across various exuberance periods. The date stamping method has detected bubbles in NIFTY 50, NIFTY NEXT 50, NIFTY 500, S&P BSE 100 whereas it also found no evidence of such exuberance in the S&P BSE SENSEX and NIFTY BANK.

The policies that are in place during exuberance periods plays an important role in the bubble formation. These are valuable information that can be utilized by the regulators and investors monitoring markets which are prone to such phenomenon. Since the model is capable of back tracking bubbles with efficacy, further extension of the paper would ideally be to enter predicted PE, PB and Dividend Yield of various sectors into the Philips et al (2015) model to simulate sample distributions. This can be a step towards predicting i.e. analyzing the current market health of the sectors under consideration.

## REFERENCES

[1] Ball, R. (2009). The Global Financial Crisis and the Efficient Market Hypothesis: What Have We Learned?. Journal of Applied Corporate Finance, 21(4), pp.8-16.

[2] Barker, R.G. (1999). Survey and Market-based Evidence of Industry-dependence in Analysts Preferences Between the Dividend Yield and Price-earnings Ratio Valuation Models. Journal of Business Finance Accounting, 26(3-4), pp.393–418.

[3] Bekaert, G. and Harvey, C. (2000). Foreign Speculators and Emerging Equity Markets. SSRN Electronic Journal.

[4] Chen, S. and Xie, Z. (2017). Detecting speculative bubbles under considerations of the sign asymmetry and size non-linearity: New international evidence. International Review of Economics & Finance, 52, pp.188-209.

[5] Dutta, A., Gahan, P. and Keshari Panda, S. (2016). Evidences of Herding Behaviour in the Indian Stock Market. Journal of Management, Vol.13.

[6] Froot, K. and Obstfeld, M. (1992). Intrinsic Bubbles. Cambridge, Mass.: National Bureau of Economic Research.

[7] Girdzijauskas, S., Štreimikiene, D., Čepinskis, J., Moskaliova, V., Jurkonyte, E. and Mackevičius, R. (2009). Formation of economic bubbles: Causes and possible preventions. Technological and Economic Development of Economy, 15(2), pp.267-280.

[8] Kim, S. and Yang, D. (2009). Do Capital Inflows Matter to Asset Prices? The Case of Korea. Asian Economic Journal, 23(3), pp.323-348.

[9] Kuepper, J. (2018). 5 Indicators that Foretold the 2008 Crash. [online] TraderHQ.com. Available at: https://traderhq.com/5-indicators-that-foretold-the-2008-crash/.

[10] Nagpal, A. and Jain, M. (2018). Efficient Market Hypothesis in Indian Stock Markets: A Re-examination of Calendar Anomalies. Amity Global Business Review.

[11] Phillips, P., Shi, S. and Yu, J. (2015). Testing For Multiple Bubbles: Historical Episodes Of Exuberance And Collapse In The S&P 500. International Economic Review, 56(4), pp.1043-1078.

[12] Sehgal, S. and Pandey, A. (2010). Equity Valuation Using Price Multiples: A Comparative Study for BRICKS. Asian Journal of Finance & Accounting, 2(1).

[13] Siddiqui, M. and Muhammad, N. (2013). Oil Price Fluctuation and Stock Market Performance - The Case of Pakistan. SSRN Electronic Journal.